\documentclass{ws-ijmpa}

\begin{document}

\markboth{MARCO PICARIELLO et al.}
{Predictions from non trivial Quark-Lepton complementarity}
%
\catchline{A}{B}{C}{D}{E}
%

\title
{Predictions from non trivial Quark-Lepton complementarity}

\author{MARCO PICARIELLO}

\address{
Dipartimento di Fisica, Universit\`a di Lecce and INFN-Lecce,
 Italia\\
\centerline{\em and}
CFTP - Departamento de F\'{\i}sica,
Instituto Superior T\'{e}cnico
 Lisboa, Portugal\\
\centerline{\ }
{\em email: Marco.Picariello@le.infn.it}}

\author{BHAG C. CHAUHAN\footnote{On leave from Govt. Degree College, Karsog (H P) India 171304.} {\ \em and}
JO\~{A}O PULIDO}

\address{ 
CFTP - Departamento de F\'{\i}sica,
Instituto Superior T\'{e}cnico
- Lisboa, Portugal
}

\author{EMILIO TORRENTE-LUJAN}

\address{ 
Dep. de Fisica, Grupo de Fisica Teorica,
Univ. de Murcia,  Murcia, Spain.}

\maketitle

\begin{history}
\received{15 June 2007}
\end{history}

\begin{abstract}
The complementarity between the quark and lepton
 mixing matrices is shown to provide robust predictions.
We obtain these predictions by first showing that the
 matrix $V_M$, product of the quark (CKM) and lepton (PMNS) mixing 
 matrices, may have a zero (1,3) entry which is 
 favored by experimental data.

We obtain that any theoretical model with a vanishing
 (1,3) entry of $V_M$ that is in agreement with quark data, solar,
 and atmospheric mixing angle leads to 
 $\theta_{13}^{PMNS}=(9{^{+1}_{-2}})^\circ$. This value is consistent
 with the present 90\% CL experimental upper limit.
We also investigate the prediction on the lepton phases.
 We show that the actual evidence, under the only assumption that the correlation matrix
 $V_M$ product of $CKM$ and $PMNS$ has a zero in the entry $(1,3)$, gives us a prediction
 for the three CP-violating invariants $J$, $S_1$, and $S_2$.
 A better determination of the lepton mixing angles will give stronger prediction
 for the CP-violating invariants in the lepton sector.
 These will be tested in the next generation experiments.
Finally we compute the effect of non diagonal neutrino
 mass in $l_i\rightarrow l_j\gamma$
 in SUSY theories with non trivial Quark-Lepton
 complementarity and a flavor symmetry.
 The Quark-Lepton complementarity and the flavor symmetry strongly
 constrain the theory and we obtain a clear prediction for the contribution
 to $\mu\rightarrow e\gamma$ and the $\tau$ decays
 $\tau\rightarrow e\gamma$ and $\tau\rightarrow \mu\gamma$.
 If the Dirac neutrino Yukawa couplings are degenerate but the
 low energy neutrino masses are not degenerate, then the
 lepton decays are related among them by the $V_M$ entries.
 On the other hand, if the Dirac neutrino Yukawa couplings are
 hierarchical or the low energy neutrino masses are degenerate,
 then the prediction for the lepton decays comes from the $CKM$
 hierarchy.

\keywords{Neutrino physics, quark lepton complementarity, Grand Unified Theories, Flavor symmetries.}
\end{abstract}

\ccode{PACS numbers: 14.60.Pq, 14.60.Lm, 96.40.Tv}

\section{Introduction}
\def\ba{\begin{eqnarray}}
\def\ea{\end{eqnarray}}
\def\beq{\begin{equation}}
\def\eeq{\end{equation}}
\def\bas{\begin{eqnarray*}}
\def\eas{\end{eqnarray*}}

\def\imag{{\rm i}}
\def\rhobar{\overline \rho}
\def\etabar{\overline \eta}
The actual experimental situation is such that we are
very close to obtain a theory of flavor that is able
to explain in a easy way all the Standard Model masses and
mixing\cite{Barbieri:1999km
}$^-$\cite{Abada:2006yd}.
The last but not least experimental ingredient have been the
neutrino data.
In fact, after the recent experimental evidences about neutrino physics
we know very well almost all the parameters in the quark
and lepton 
sectors.
 We measured all the quark and charged lepton masses,
and the value of the difference between the square of the neutrino masses
$\Delta m_{12}^2=m_1^2-m_2^2$ and $|\Delta m_{23}^2|=|m_3^2-m_2^2|$. 
We also know the value of the quark mixing angles and phases, and the two mixing angles
$\theta_{12}$ and $\theta_{23}$ in the lepton sector.
The challenger for the next future
\cite{deHolanda:2004fd
}$^-$\cite{Balantekin:2004hi}
 will be to determine the sign of $\delta m^2_{23}$
(i.e. the hierarchy in the neutrino sector), the absolute scale of the neutrino masses,
and the value of the 3rd lepton mixing angle $\theta_{13}$
(in particular if is it zero or not).
Finally, if $\theta_{13}$ is not too small, there is a hope to measure the CP
violating phases.

From all these results we are able to extract strong
constraints on the flavor structure of the SM.
In particular the neutrino data were determinant to
clarify the role of the discrete symmetry in flavor physics.
The neutrino experiments confirm \cite{Fogli:2004zn
}$^-$\cite{Aguilar-Saavedra:2003yc}
that the Pontecorvo-Maki-Nakagawa-Sakata
(PMNS) \cite{Pontecorvo:1967fh,Maki:1962mu}
lepton mixing matrix $U_{PMNS}$ contains large mixing angles.
For example the atmospheric mixing $\theta_{23}^{PMNS}$
is compatible with $45^\circ$
and the solar mixing $\theta_{12}^{PMNS}$ is $\approx 34^\circ$ 
\cite{Altarelli:2004za
}$^-$\cite{Pascoli:2003ke}.
These results should be compared with the third lepton mixing angle
$\theta_{13}^{PMNS}$ which is very small and even compatible with zero 
\cite{Palomares-Ruiz:2004tk,Apollonio:1999ae}, and with the quark mixing angles 
in the CKM matrix~\cite{Cabibbo:1963yz,Kobayashi:1973fv}.
The disparity that nature indicates between quark and lepton mixing
angles has been viewed in terms of a 'Quark-Lepton complementarity' (QLC)
\cite{Georgi:1979df
}$^-$\cite{Agarwalla:2006dj}
which can be expressed in the relations
\begin{eqnarray}\label{eq:QLCnaive}
\theta_{12}^{PMNS}+\theta_{12}^{CKM}\simeq 45^\circ\,;
\quad\quad
 \theta_{23}^{PMNS}+\theta_{23}^{CKM}\simeq 45^\circ\,.
\end{eqnarray}
Possible consequences
of QLC have been investigated in the literature
and in particular a simple correspondence between the $U_{PMNS}$ and $U_{CKM}$
matrices has been proposed~\cite{Ferrandis:2004mq
}$^-$\cite{Ma:2005qy}
and analyzed in terms of a 
correlation matrix~\cite{Xing:2005ur
}$^-$\cite{Picariello:2007yn}.
The relations in eq. (\ref{eq:QLCnaive}) are related to the 
parametrization used for the CKM and PMNS mixing
matrix.
From a more general point of view, we can define a correlation matrix $V_M$
as the product of the PMNS and CKM mixing matrices,
$V_M=U_{CKM}\, U_{PMNS}\,.$
A lot of efforts have been done to obtain the
{\em most favorite} pattern for the matrix
$V_M$~\cite{Georgi:1979df,Frampton:2004vw
}$^-$\cite{Smirnov:2006qj}.
The naive QLC relations in eq. (\ref{eq:QLCnaive}) seems to
implies $V_M$ to be Bi-Maximal, i.e. in the standard parametrization
it contains two maximal mixing angle, and a third angle to be zero.
Despite the naive relations between the PMNS and CKM angles, 
a detailed analysis shows that the correlation matrix $V_M=U_{CKM}U_{PMNS}$
is phenomenologically compatible with a TriBi-Maximal pattern, 
and only marginally with a Bi-Maximal pattern.
From actual experimental evidences a non trivial Quark-Lepton
complementarity arises \cite{Chauhan:2006im}, i.e. we learn that
$V_M$ Bi-Maximal, although it is not ruled out
by the experiments, is excluded at 90\% CL in non SUSY models,
or in SUSY models with $\tan \beta<40$ where the RGE correction are
negligible \cite{Schmidt:2006rb
}$^-$\cite{Antusch:2005gp},
and a non trivial Quark-Lepton
complementarity arises \cite{Chauhan:2006im}.
Future experiments on neutrino physics, and in particolar
in the determination of $\theta_{23}$ and the CP violating parameter
$J$, will be able to better clarify if a trivial Quark-Lepton
complementarity, i.e. $V_M$ Bi-Maximal, is ruled out in favor of
a non trivial Quark-Lepton complementarity,
i.e. $V_M$ TriBi-Maximal or even more structured \cite{Picariello:2006sp}.
Unitarity then implies $U_{PMNS}=U^{\dagger}_{CKM}V_M$ 
and one may ask where do the large lepton mixings come from?
Is this information implicit in the form of the $V_M$ matrix?
This question has been widely investigated in the literature,
but its answer is still open.
However the fact $V_M$ 
has a clear non trivial structure and the strong indication of gauge
coupling unification allow us to obtain in a straightforward way
constraints on the high energy spectrum too.
Within this framework we get some informations
about flavor physics from the correlation matrix $V_M$ itself.
It is very impressive that for some discrete flavor symmetries
such as $A_4$ dynamically broken into $Z_3$, 
as in Refs. \refcite{Morisi:2007ft} and \refcite{Altarelli:2005yp},
or $S_3$ softly broken into $S_2$, 
as in Ref. \refcite{Morisi:2005fy}, the TriBi-Maximal structure
appears in a natural way.
In fact in some Grand Unification Theories (GUTs) the direct 
QLC correlation between the $CKM$ and the $PMNS$ mixing matrix can
be obtained. In this class of models, the $V_M$ matrix is determined
by the heavy Majorana neutrino mass
matrix~\cite{Georgi:1979df,Antusch:2005ca}.
Moreover as long as quarks and leptons are inserted in the same 
representation of the underlying gauge group,
we need to include in our definition of $V_M$ arbitrary
but non trivial phases between the quark and lepton matrices. 
Hence we will generalize the relation $V_M=U_{CKM}\cdot  U_{PMNS}$
to 
\begin{equation}\label{eq:fund}
V_M=U_{CKM}\cdot \Omega \cdot U_{PMNS}\,
\end{equation}
where the quantity $\Omega$ is a diagonal matrix
$\Omega={\rm diag}(e^{\imag\omega_i})$ and the three phases
$\omega_i$ are free parameters (in the sense that they are
not restricted by present experimental evidence).

In this paper we will show how the investigation of
the correlation matrix $V_M$ based on eq.~(\ref{eq:fund})
implies that there is a zero texture of $V_M$, namely
$V_{M_{13}}=0$.
The conclusion for matrix $V_M$ is that the correlation
between the matrices $U_{CKM}$ and  $U_{PMNS}$ is
rather nontrivial. Then, by using this fact we will
report the predictions that can be obtained from
experimental data and QLC for $\theta_{13}^{PMNS}$,
CP violating parameters in the lepton sector and
the lepton number violating decays.
The plan of the work is the following.
In section~{\bf\ref{sec:VM}} we study the numerical ranges of
$V_M$ entries with the aid of a Monte Carlo simulation and we
will show that the vanishing of the $(1,3)$ entry 
is favored by the data analysis.
After that we present the matter from a different
point of view:
we start from a zero $(1,3)$ $V_M$ entry (e.g. a Bi-Maximal or 
TriBi-Maximal matrix) and we derive the consequent predictions.
In section~{\bf\ref{sec:PMNS}} we get a small value for
$\theta_{13}^{PMNS}$ with  a sharp prediction
\begin{equation}\label{eq:prediction}
\theta_{13}^{PMNS}=(9\pm^{1}_{2})^{\circ}\,,
\end{equation} 
for the $U_{PMNS}$ 
lepton mixing angle through 
\beq
U_{PMNS}=(U_{CKM}\cdot  \Omega)^{-1}\cdot  V_M
\eeq
In sec. {\bf\ref{sec:CP}}, with the aid of the Monte Carlo simulation,
we study the numerical correlations of the lepton
CP violating phases $J$, $S_1$, and $S_2$ with respect to the
mixing angle $\theta_{12}^{PMNS}$.
In Sec. {\bf\ref{sec:muegamma}} we compute the value of
the contribution to the $l_i\rightarrow l_j\gamma$ processes from
a non diagonal Dirac neutrino Yukawa coupling.
By using the non trivial Quark-Lepton complementarity and
the see-saw mechanism we will compute the explicit 
spectrum of the heavy neutrino. This will allow us to
investigate the relevance of the form of $V_M$ in the
$l_i\rightarrow l_j\gamma$.

\section{Which $V_M$ does the phenomenology imply?}\label{sec:VM}
In this section we investigate the value of the $V_M$ matrix
entries concentrating in particular in the (1,3) entry and the important 
mixing angle $\theta_{13}^{V_{M}}$ to which it is directly related. We then
explicitly study the allowed values of the $V_M$ angles, finally concluding 
that $\sin^2 \theta_{13}^{V_{M}}=0$ is the value most favored by the data.
We will be using the Wolfenstein parameterization \cite{Wolfenstein:1983yz}
of the $U_{CKM}$ matrix in its unitary form \cite{Buras:1994ec}
where one has the relation 
\beq\label{CKM}
\sin \theta_{12}^{CKM}=\lambda
\quad\quad
\sin \theta_{23}^{CKM}=A\lambda^2
\quad\quad
\sin \theta_{13}^{CKM}e^{-\imag \delta^{CKM}}=A \lambda^3(\rho-\imag \eta)
\eeq
to all orders in $\lambda$.
The lepton mixing matrix $U_{PMNS}$ is parameterized as
\ba
U_{PMNS}=
 U_{23}\cdot \Phi\cdot U_{13}\cdot\Phi^\dagger\cdot U_{12}\cdot \Phi_m.
\ea
Here $\Phi$ and $\Phi_m$ are diagonal matrices containing the Dirac and
Majorana CP violating phases, respectively
$\Phi={\rm diag}(1, 1, e^{\imag \phi})$
and $\Phi_m={\rm diag}(e^{\imag \phi_1}, e^{\imag \phi_2}, 1)$, so that
\ba\label{PMNS}
U_{PMNS}=\left(
\begin{matrix}
e^{\imag\,\phi_1} c_{12}\,c_{13} &
            e^{\imag\,\phi_2} c_{13}\,s_{12} &
                   s_{13}e^{-\imag \,\phi}\cr
 e^{\imag\,\phi_1}\left(
    -c_{23}\,s_{12}-e^{\imag \,\phi}\,c_{12}\,s_{13}\,s_{23}\right)&
e^{\imag\,\phi_2}\left(
            c_{12}\,c_{23} - e^{\imag \,\phi}\,s_{12}\,s_{13}\,s_{23}\right)&
                   c_{13}\,s_{23}\cr
 e^{\imag\,\phi_1}\left(
   - e^{\imag \,\phi}\,c_{12}\,c_{23}\,s_{13} + s_{12}\,s_{23}\right)&
e^{\imag\,\phi_2}\left(
            - e^{\imag \,\phi}\,c_{23}\,s_{12}\,s_{13}-c_{12}\,s_{23}\right)&
                   c_{13}\,c_{23}\,
\end{matrix}\right)\nonumber
\ea
The investigation we perform for the $V_M$ matrix starts from the 
fundamental equation $V_M=U_{CKM}\cdot \Omega \cdot U_{PMNS}$ and
uses the experimental ranges and constraints on lepton mixing angles. 
We resort to a Monte Carlo simulation with two-sided Gaussian distributions
around the mean values of the observables. 
We use the updated values for the $CKM$ and $PMNS$ mixing matrix,
given at $95\%$CL by \cite{Charles:2004jd}
\ba\label{bestfit}
&&\begin{tabular}{cc}
$\lambda = 0.2265^{+0.0040}_{-0.0041}$
\,,&
$A=0.801^{+0.066}_{-0.041}$
\,,\cr\cr
$\etabar = 0.189^{+0.182}_{-0.114}$
\,,&
$\rhobar = 0.358^{+0.086}_{-0.085}$
\,,
\end{tabular}
\\
\mbox{with\quad\quad}&&
\rho + i\eta =
\frac{\sqrt{1-A^2\lambda^4}(\rhobar+i\etabar)}{\sqrt{1-\lambda^2}
\left[1 - A^2\lambda^4(\rhobar+i\etabar)\right]}\,;
\end{eqnarray}
and~\footnote{The lower uncertainty for $\sin^2\theta_{13}$ is purely formal,
and correspond to the positivity constraint $\sin^2\theta_{13}\geq 0$.} 
\cite{Altarelli:2004za
}$^-$\cite{Pascoli:2003ke}
\ba\label{bestfit2}
\begin{tabular}{cc}
$\sin^2\theta_{23}^{PMNS} = 0.44\times\left(1^{+0.41}_{-0.22}\right)$
\,,&
$\sin^2\theta_{12}^{PMNS} = 0.314\times\left(1^{+0.18}_{-0.15}\right)$
\,,\cr\cr
$\sin^2\theta_{13}^{PMNS} = \left(0.9^{+2.3}_{-0.9}\right)\times 10^{-2}$
\,.
\end{tabular}
\ea
With the aid of a Monte Carlo program we generated the values for
each variable: for the sine square of the lepton mixing angles and
for the quark parameters $A,~\lambda,~\bar\rho,~\bar\eta$ we took
two sided Gaussian distributions with central values and standard
deviations taken from eqs. (\ref{bestfit}-\ref{bestfit2}). For the unknown phases we
took flat random distributions in the interval $[0,2\pi]$. We divided
each variable range into short bins and counted the number of occurrences
in each bin for all the variables, having run the program $10^6$ times.
In this way the corresponding histogram is smooth and the number of
occurrences in each bin is identified with the probability density
at that particular value. A comparatively high value of this
probability density extending over a wide range in
the variable domain means a high probability for the variable to lie
in this range, therefore that such range is 'favored' by the
data being used as Monte Carlo input. Conversely higher probability
implies better compatibility with experimental data, while lower
probability means poor or no compatibility with data.

\begin{figure}[hbt]
\centering
{\epsfig{file=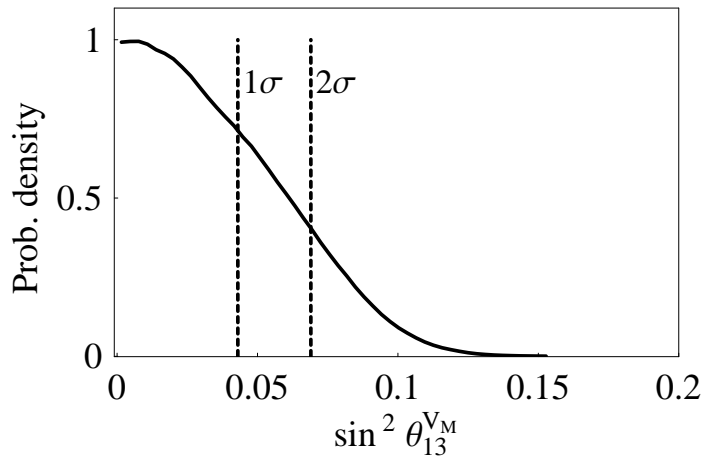,height=6.cm}}
\caption{The distribution, normalized to one at the maximum,
of $\sin^2\theta_{13}^{V_M}$
 obtained from the definition of the correlation mixing matrix $V_M$
given in eq.~(\ref{eq:fund}) by using a Monte Carlo simulation
of all the experimental data.
We also plot the $1\sigma$ and the $2\sigma$ lines.}
\label{fig:f3}
\end{figure}
The range of $\tan^2\theta_{23}^{V_M}$ which is compatible
with experiments at $90\%$CL is the interval $[0.35, 1.4]$, so that 
$\tan^2\theta_{23}^{V_M}=1.0$ is consistent with data.
For $\tan^2\theta_{12}^{V_M}$ we obtain a range between
$0.25$ and $1.1$ at $90\%$CL and so $\tan^2\theta_{12}^{V_M}=1.0$ 
(which corresponds to a Bi-Maximal matrix) only
within 3$\sigma$.
Moreover the value $\tan^2\theta_{12}^{V_M}=0.5$ (which corresponds
to a TriBi-Maximal matrix), is well inside the allowed range.
We checked that for $\tan^2\theta_{12}^{V_M}=0.3,~{\rm and}~0.5$ 
the resulting distribution for
$\tan^2\theta_{12}^{PMNS}$ is compatible with the experimental data. 
Instead maximal $\theta_{12}^{V_M}$ and $\theta_{23}^{V_M}$ taken together
are disfavored, as the solar angle is hardly compatible
with the corresponding allowed interval.
We also checked that the distribution of $\tan^2\theta_{23}^{PMNS}$
for $\tan^2\theta_{23}^{V_M}\in\{0.5,1.0,1.4\}$ with
$\tan^2\theta_{12}^{V_M}=0.5$ are compatible with the
experimental data.

In fig.\ref{fig:f3} we plot the distribution for
$\sin^2\theta_{13}^{V_M}$.
We see that $\sin^2\theta_{13}^{V_M}=0$
is not only allowed by the experimental data,
 but also it is the preferred value.
In the next section we will see that this has important consequences
in the model building of flavor physics.

\section{Prediction for $\theta_{13}^{PMNS}$}\label{sec:PMNS}

In this section
we investigate the consequences of a $V_M$ correlation matrix
with zero (1,3) entry on the still experimentally undetermined
$\theta_{13}^{PMNS}$ mixing angle.
In particular we will see that the $\theta_{13}^{PMNS}$
prediction arising from eq.~(\ref{eq:fund}) or, equivalently,
\ba\label{PMNSexp}
U_{PMNS}&=&(U_{CKM}\cdot \Omega)^{-1}\cdot V_M
\ea
is quite stable against variations in the form of $V_M$ allowed
by the data.

As previously shown (see section~{\bf\ref{sec:VM}}),
the data favors a vanishing (1,3) entry in $V_M$.
So in the whole following analysis we fix
$\sin^2\theta_{13}^{V_M}=0$. We allow the 
$U_{CKM}$ parameters to vary, with a two-sided Gaussian distribution,
within the experimental ranges given in eq.~(\ref{bestfit}), while for the 
$\Omega$ phases in eq. (\ref{PMNSexp}) we take flat distributions in the interval
$[0,2\pi]$.
We make Monte Carlo simulations for different values of $\theta_{12}^{V_M}$ 
and $\theta_{23}^{V_M}$ mixing angles, allowing
$\tan^2\theta_{12}^{V_M}$ and $\tan^2\theta_{23}^{V_M}$ 
to vary respectively within the intervals $[0.3,1.0]$ and $[0.5,1.4]$ in 
consistency with the lepton and quark mixing angles
(see section~{\bf\ref{sec:VM}}).

From eq.~(\ref{PMNSexp}), the parameterization of the CKM mixing matrix
in eq.~(\ref{CKM}) and the definition of the phase matrix $\Omega$ 
we get  
\ba
(U_{PMNS})_{13} &=&
e^{-i\omega_1}\Big[
\left(1-\frac{\lambda^2}{2}\right)\sin \theta_{13}^{V_M}e^{-i \phi^{V_M}}
-\lambda \sin \theta_{23}^{V_M} \cos \theta_{13}^{V_M}
\nonumber\\&&\quad\quad\quad
 + A\lambda^3(-\rho+i\,\eta+1)\cos \theta_{23}^{V_M} \cos \theta_{13}^{V_M}
+ O(\lambda^4)
 \Big]\,,
\ea
so that
\ba\label{th13PMNS_A}
\sin^2\theta_{13}^{PMNS} &=&
\sin^2\theta_{23}^{V_M}\lambda^2 + O(\lambda^3)\,,
\ea
where we have used the fact that $\sin^2\theta_{13}^{V_M}=0$
and $A\approx O(1)$.
We see that $\sin^2\theta_{13}^{PMNS}$ does
 not depend on $\tan^2\theta_{12}^{V_M}$.
For this reason the parameter $\sin^2\theta_{13}^{PMNS}$ needs to be studied
as a function of $\tan^2\theta_{23}^{V_M}$ only. Fixing for definiteness
$\tan^2\theta_{12}^{V_M}=0.5$ and taking the three different values
$\tan^2\theta_{23}^{V_M}\in\{0.5, 1.0, 1.4\}$, we computed  
the corresponding distributions of $\sin^2\theta_{13}^{PMNS}$. 
We note that these values of $\tan^2\theta_{23}^{V_M}$ practically 
cover the whole range consistent with the data.
It is seen that the $\sin^2\theta_{13}^{PMNS}$ distributions are
quite sharply peaked around maxima of $7.3^\circ$, $8.9^\circ$ and 
$9.8^\circ$. Recalling that the shift of this maximum is effectively
determined 
by the parameter $\tan^2\theta_{23}^{V_M}$ which was chosen to span most
of its physically allowed range, it is clear that we have a stable 
prediction for $\theta_{13}^{PMNS}$.

\begin{figure}[hbt]
\centering
{\epsfig{file=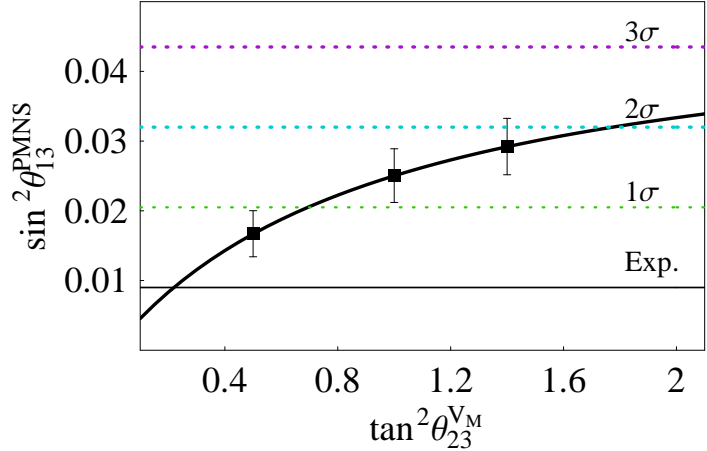,height=6.cm}} 
\caption{The allowed values for $\sin^2\theta_{13}^{PMNS}$ as a function of 
$\tan^2\theta_{23}^{V_M}$ under the assumption that
 $\sin^2\theta_{13}^{V_M}=0$.
We report the central and $3 \sigma$ values,
and the approximate analytical dependence given in eq.~(\ref{th13PMNS_A}).
We also plot the experimental central value, the $1\sigma$, the $2\sigma$,
and the $3\sigma$. 
We fixed $\tan^2\theta_{12}^{V_M}=0.5$ for definiteness.}
\label{fig:f5}
\end{figure}
In order to better clarify this stability, we show 
in fig.~\ref{fig:f5} the mean and the standard deviation of 
$\sin^2\theta_{13}^{PMNS}$ obtained with our Monte Carlo simulation for the 
three chosen values of $\tan^2\theta_{23}^{V_M}$.
In addition we plot the analytic
dependence of $\sin^2\theta_{13}^{PMNS}$ given by eq.~(\ref{th13PMNS_A}) with 
the central value of $\lambda$, the best fit point of
$\sin^2\theta_{13}^{PMNS}$ and its 1$\sigma$, 2$\sigma$
and 3$\sigma$ from standard analysis. 
Our prediction for $\theta_{13}^{PMNS}$ then follows from the experimental data on
$\lambda$ ,$A$, $\rho$, $\eta$, $\tan^2\theta_{12}^{PMNS}$
and $\tan^2\theta_{23}^{PMNS}$ and the values of $\tan^2\theta_{12}^{V_M}$,
$\tan^2\theta_{23}^{V_M}$ are taken in the intervals $[0.3,1.0]$, $[0.5,1.4]$ 
respectively, as allowed by the data. For a vanishing $(1,3)$ entry of the matrix 
$V_M$ we finally find $\theta_{13}^{PMNS}$ in the interval $[7^\circ,10^\circ]$.  

To conclude this section we note that another prediction for a small 
$\theta_{13}^{PMNS}$ has recently been derived \cite{Harada:2005km} 
\begin{equation}
\theta_{13}^{PMNS}=9^{\circ}+O(sin^{3}\theta_{12}^{CKM}).
\end{equation}
This follows from an assumed Bi-maximality of a matrix relating
Dirac to Majorana neutrino states together with the assumption 
that neutrino mixing is described by the CKM matrix at the grand unification
scale. Our approach on the other hand is free from any {\it ad hoc} assumptions. 
We show that it is a zero texture of the $V_M$
correlation matrix, namely $V_{M_{13}}=0$, together with all the experimental values
of the quark and lepton mixing angles, that predicts
$\theta_{13}^{PMNS}=(9 \pm^{1}_{2})^{\circ}$. More importantly, in sec. {\bf\ref{sec:VM}}
we show that the vanishing of this entry is favored by the data. 
Condition $V_{M_{13}}=0$ is compatible with $V_M$ being 
Bi-Maximal (i.e. with two angles of $45^\circ$ and a
vanishing one), TriBi-Maximal (i.e. with one angle of $45^\circ$, one with
$\tan^2\theta=0.5$ and a third vanishing one) or of any other form.
Furthermore we make use of a phase matrix $\Omega$, see 
eq.~(\ref{eq:fund}), that takes account of the mismatch between 
the quark and lepton phases
and consider Majorana phases in the $U_{PMNS}$ matrix with a flat random 
distribution. 

\section{CP violating invariants in the lepton sector}\label{sec:CP}
In this section we investigate the consequences of a $V_M$
correlation matrix with a zero $(1,3)$
entry on the undetermined CP violating parameters in the lepton sector.
There are two kind of invariants parameterizing CP violating effect.
The Jarlskog invariant $J$ that parametrizes the effects related 
to the Dirac phase,
and the two invariants $S_1$ and $S_2$ that parametrize the effects related
to the Majorana phases.
The $J$ invariant describes all CP breaking
observables in neutrino oscillations.
It is the equivalent of the Jarlskog invariant in the quark sector.
It is given by
\begin{equation}\label{eq:JU}
J=Im\{U_{\nu_e \nu_1}U_{\nu_\mu \nu_2}U_{\nu_e \nu_2}^* U_{\nu_\mu \nu_1}^*\}\,.
\end{equation}
In the parametrization of eq. (\ref{PMNS}) one has
\begin{eqnarray}\label{eq:J}
J&=&\frac{1}{8}\sin2\theta_{12}\ \sin2\theta_{23}\ \sin2\theta_{13}\ \cos\theta_{13}\ \sin\phi\,.
\end{eqnarray}
Then we have the two invariants $S_1$ and $S_2$ 
that are related to the Majorana phases.
They are
\begin{eqnarray}\label{eq:SU}
S_1 &=& Im\{U_{\nu_e \nu_1} U_{\nu_e \nu_3}^*\}\\
S_2 &=& Im\{U_{\nu_e \nu_2} U_{\nu_e \nu_3}^*\}\nonumber
\end{eqnarray}
In the parametrization of eq. (\ref{PMNS}) we have
\begin{eqnarray}\label{eq:S}
S_1 &=& \frac{1}{2} \cos \theta_{12} \sin 2\theta_{13} \sin(\phi+\phi_1)\\
S_2 &=& \frac{1}{2} \sin \theta_{12} \sin 2\theta_{13} \sin(\phi+\phi_2)\nonumber
\end{eqnarray}
The two Majorana phases appear in $S_1$ and $S_2$ but not in $J$.

As show in sec. {\bf\ref{sec:VM}}, the data favors a
vanishing $(1,3)$ entry in the correlation matrix $V_M$ \cite{Chauhan:2006im}.
So in the whole analysis we fix $\sin^2 \theta_{13}^{V_M}=0$.
Moreover $\tan^2\theta_{12}^{V_M}$ and $\tan^2\theta_{23}^{V_M}$ are allowed
to vary respectively within the intervals $[0.3,1.0]$ and $[0.5,1.4]$.
We allow the $U^{CKM}$ parameters to vary, with a two-sided Gaussian distribution,
within the experimental ranges given in eq. (\ref{bestfit}).
For the $\Omega$ phases in eq. (\ref{eq:fund}) we take flat distributions
in the interval $[0,2\pi]$.
We make Monte Carlo simulations for different values of $\theta_{12}^{V_M}$
and $\theta_{23}^{V_M}$ mixing angles, allowing $\tan^2\theta_{12}^{V_M}$
and $\tan^2\theta_{23}^{V_M}$ to vary respectively within their allowed intervals,
in consistency with the lepton and quark mixing angles.
From eq. (\ref{eq:J}), by using the fact that $\theta_{13}$ is small and that
$\theta_{23}$ is maximal, we get
$$
J\approx \frac{1}{8} \sin 2\theta_{12} \sin 2\theta_{13} \sin \phi
$$
This expression tells us that the $J$ parameter is within the range $|J|<0.042$.
However there is a non trivial correlation between $J$ and $\theta_{12}^{PMNS}$.
Because the $CKM$ is given by the experimental data,
and $(V_M)_{13}$ is fixed to be zero, the phase $\phi$ and the $\theta_{13}^{PMNS}$
angle are almost fixed as a function of $\theta_{12}^{PMNS}$.

\begin{figure}[hbt]
\centering
{\epsfig{file=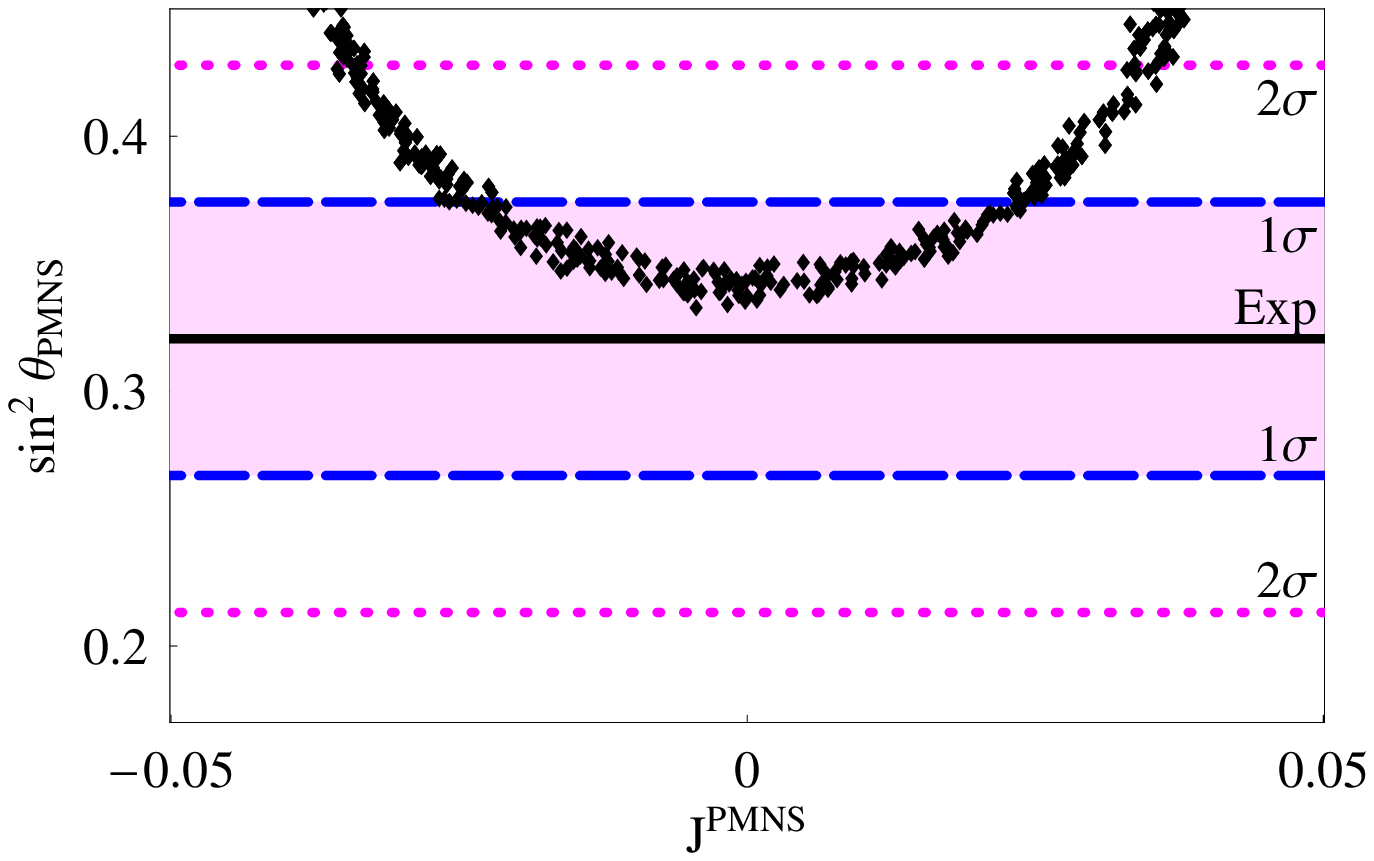,height=3.5cm}}
\vskip -0.5truecm
\caption{The correlation between the Dirac CP violating 
parameter $J$ and $\sin^2\theta_{12}^{PMNS}$ for $V_M$
Bi-Maximal.
We also plot the experimental central value, the
$1\sigma$, and the $2\sigma$ for
$\sin^2 \theta_{12}^{PMNS}$.
}
\label{fig:f11}
%
\vskip 0.5truecm
\centering
{\epsfig{file=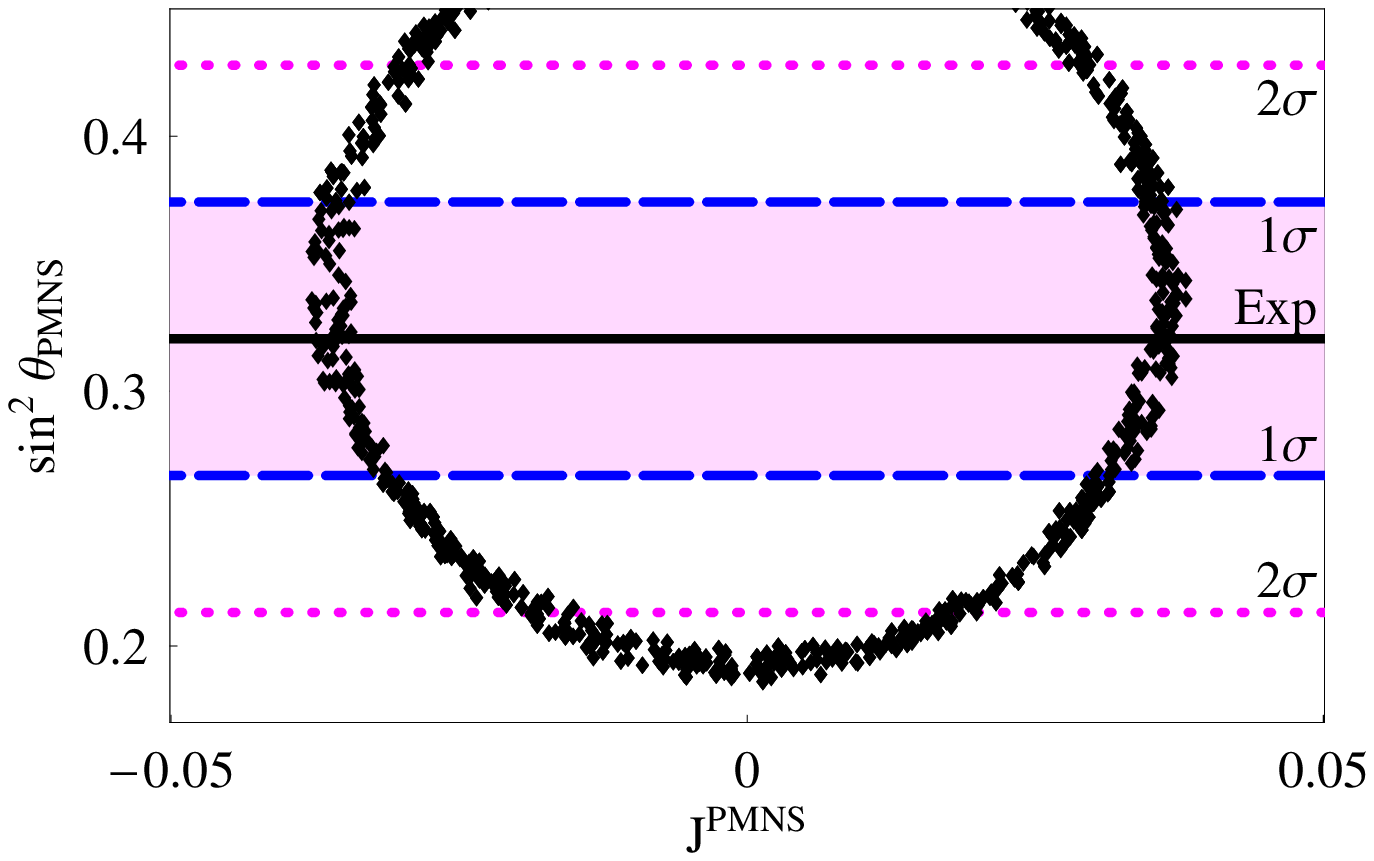,height=3.5cm}}
\vskip -0.5truecm
\caption{The correlation between the Dirac CP violating 
parameter $J$ and $\sin^2\theta_{12}^{PMNS}$ for $V_M$ TriBi-Maximal.
}
\label{fig:f12}
%
\vskip 0.5truecm
\centering
{\epsfig{file=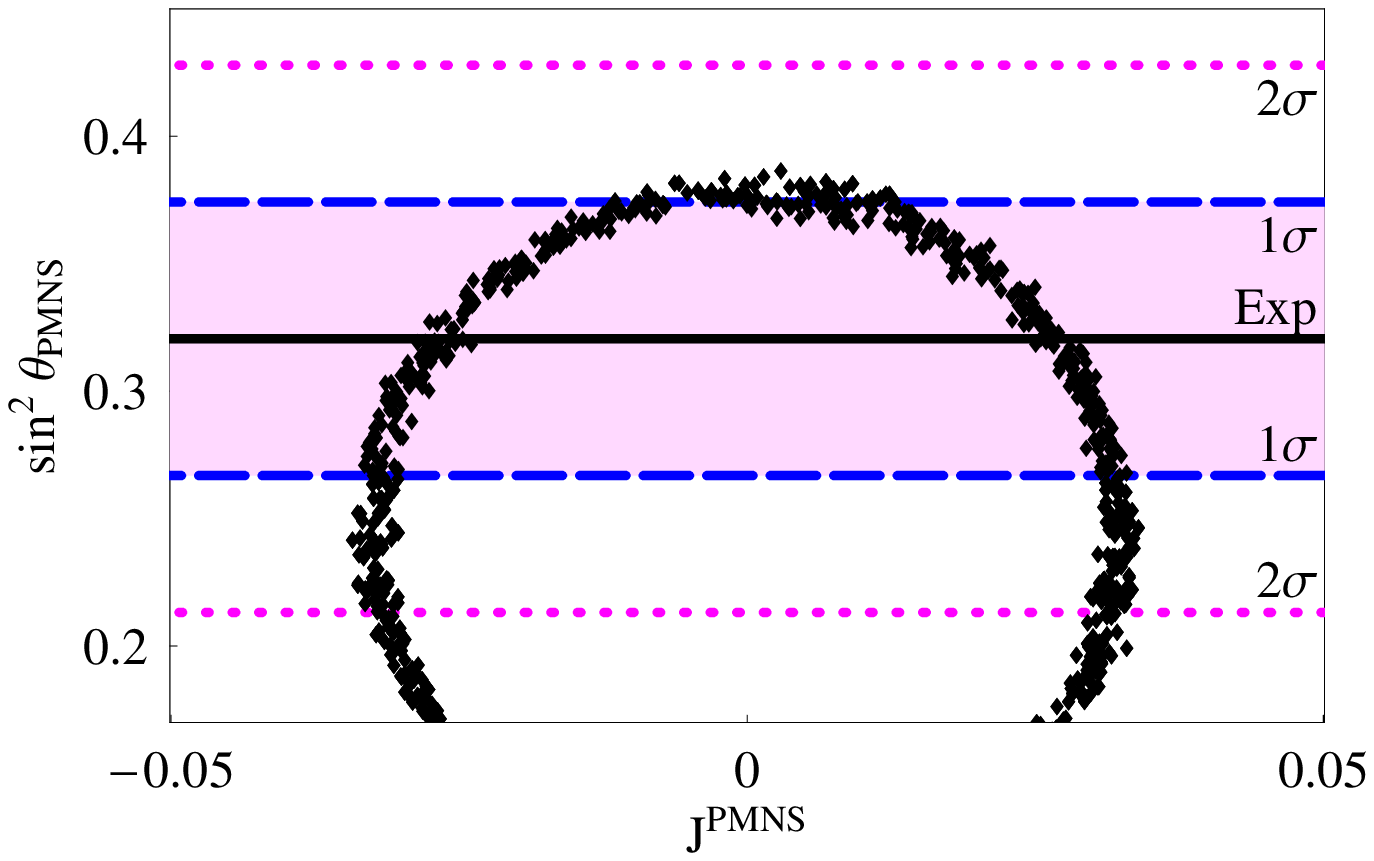,height=3.5cm}}
\vskip -0.5truecm
\caption{The correlation between the Dirac CP violating 
parameter $J$ and $\sin^2\theta_{12}^{PMNS}$
for $V_M$ such that $\tan^2\theta_{12}^{V_M}=0.4$.
}
\label{fig:f13}
\end{figure}
In figs. \ref{fig:f11}-\ref{fig:f13}
we report the result of our simulation for $J$.
We plot the correlation between the $J$ invariant and $\sin^2\theta_{12}^{PMNS}$ for $V_M$ Bi-Maximal (fig. \ref{fig:f11}),
TriBi-Maximal (fig. \ref{fig:f12}),
and $V_M$ with $\tan^2\theta_{12}^{V_M}=0.4$ (fig. \ref{fig:f13}).
First of all, from fig. \ref{fig:f11}, we see that the {\em solar} mixing angle
$\theta_{12}^{PMNS}$ is constrained to have $\sin^2 \theta_{12}^{PMNS}>0.36$
for $V_M$ Bi-Maximal.
From figs. \ref{fig:f11}-\ref{fig:f13} we see the correlation between the
structure of $V_M$ and the CP violating invariant $J$. In particular,
for $V_M$ Bi-Maximal $J$ is close to zero. For $V_M$ TriBi-Maximal
$|J|$ is around its maximum value $0.042$. Finally for $V_M$ such that
$\tan^2\theta_{12}^{V_M}=0.4$ we get that $J$ can be any value between
$-0.04$ and $0.04$. 
We also see that a better determination of the
$\sin^2\theta_{12}^{PMNS}$ could give a stronger
prediction for the $J$ invariant in the case of $V_M$ TriBi-maximal.

Similar results hold
for $S_1$ and $S_2$ (the plots have similar shapes).
The expressions in eqs. (\ref{eq:S}) give us the range for these invariants:
\begin{eqnarray}
|S_1|<0.14&\quad\quad&|S_2|<0.11
\end{eqnarray}
We obtain that for $V_M$ Bi-Maximal the Majorana CP 
invariant $S_1$ is close to zero, for $V_M$ TriBi-Maximal $S_1$ is around
$0.13$. Finally for $V_M$ such that $\tan^2\theta_{12}^{V_M}=0.4$
we obtain that $S_1$ can be any value between $-0.14$ and $0.14$.
We see that also in this case a better determination of the
$\theta_{12}^{PMNS}$ mixing angle will give a stronger constraint for the
$S_1$ and $S_2$ invariant for $V_M$ TriBi-Maximal.
As for $J$, these correlations of $S_1$ and $S_2$ with respect to $\theta_{12}^{PMNS}$
are predictions of any theoretical model that gives a relation of the type
$V_M=U^{CKM}\, \Omega\,  U^{PMNS}$ with $(V_M)_{13}=0$.
In the next section we will show how to construct an explicit model that predict
$(V_M)_{13}=0$.

\section{$l_i\rightarrow l_j\gamma$}\label{sec:muegamma}
In this section we compute the effect of non diagonal neutrino
mass on $l_i\rightarrow l_j\gamma$ 
in SUSY theories with non trivial Quark-Lepton
complementarity an a flavor symmetry.
The correlation matrix $V_M=U_{CKM}U_{PMNS}$ is such that its
$(1,3)$ entry, as preferred by the actual experimental data, is zero.
We obtain a clear prediction for the contribution to
$l_i\rightarrow l_j\gamma$.
%
%
There are three cases. They depend on the spectrum of
the Dirac neutrino mass matrix and the low energy neutrino.
We may have:
1) hierarchical Dirac neutrino eigenvalues (in this case we
have very hierarchical right-handed neutrino masses);
2) degenerate Dirac neutrino eigenvalues, with non degenerate
low energy neutrino masses (in this case the
hierarchy of the right-handed neutrino masses is close to the
one of the low energy spectrum);
3) degenerate Dirac neutrino eigenvalues and low energy neutrino
spectrum (that implies right-handed neutrino close to be degenerate).
For each of these cases we have different contributions to
$l_i\rightarrow l_j\gamma$.
We will show that only when Dirac
neutrino eigenvalues are degenerate and 
low energy neutrino masses are not degenerate,
then the explicit form of $V_M$ plays an important role.

The contribution at first order approximation
to the process $l_i\rightarrow l_j\gamma$ in SUSY
models is given by
\begin{eqnarray}\label{eq:BR}
BR(l_i\rightarrow l_j\gamma)
     &\propto&
\frac{\Gamma(l_i\rightarrow e\nu\nu)}{\Gamma(l_i)}
\frac{\alpha^3}{G_f m_s^8 v_u^4}\tan^2\beta
\left(\frac{3m_0+A_0}{8\pi^2}\right)^2
\left|\left(\tilde M_D L \tilde M_D^\dagger\right)_{ij}\right|^2
\nonumber
\end{eqnarray}
where $m_0$ is the universal scalar mass,
$A_0$ is the universal trilinear coupling parameter,
$\tan\beta$ is the ratio of the vacuum
expectation values of the up and down Higgs doublets,
and $m_s$ is a typical mass of superparticles
with \cite{Petcov:2003zb}
$ m^8_s\approx 0.5 m_0^2 M^2_{1/2}(m^2_0+ 0.6 M^2_{1/2})^2$,
where $M_{1/2}$ is the gaugino mass.
The matrix $L_{ij}={\bf 1}_{ij}\log M_x/M_i$ takes into account the
RGE effects on the Majorana right-handed neutrino masses.
In fact the eq. (\ref{eq:BR}) is computed in the base where the
Yukawa of the charged lepton and the Majorana neutrino mass
are diagonal.
Eq. (\ref{eq:BR}) is valid in the base where right-handed Majorana
neutrino mass matrix, charged lepton mass matrix and weak
gauge interactions are diagonal.
The experimental limit for the branching ratio of
$\mu\rightarrow e\gamma$ is $1.2\times 10^{-11}$ at $90\%$ of confidence
level \cite{Brooks:1999pu} and it could go
down to $10^{-14}$ as proposed by MEG collaboration.

In supergravity theories if the effective
Lagrangian is defined at a scale higher than the
Grand Unification scale, then the matter fields
have to respect the underlying gauge and flavor symmetry.
Hence, we expect quark-lepton correlations among entries of 
the sfermion mass matrices. In other words, the quark-lepton
unification seeps also into the SUSY
breaking soft sector \cite{Ciuchini:2007ha}.
In general we do not get strongly renormalization effects
on flavor violating quantities from the heavy neutrino
scale to the electroweak scale because the absence 
of flavor violation. In fact the remaining flavor
violation related to the low energy neutrino sector
gives negligible contribution with the exception of
the case with high degenerated neutrino and $\tan\beta>40$
\cite{Antusch:2005gp,Schmidt:2006rb}.

Let be $M_R$ the Majorana mass matrix for the right neutrino
and $M_{D}$ the Dirac mass matrix.
Under the assumption that the low energy neutrino masses are given by
the see-saw of Type I we have that the light neutrino mass matrix
is given by
\begin{equation}\label{eq:Mnu}
M_\nu = M_{D} \frac{1}{M_R} M_{D}^T\,.
\end{equation}
The lepton mixing matrix is
\begin{eqnarray}\label{eq:PMNS}
U_{PMNS} = U_l^\dagger U_\nu = U_l^\dagger U_0 V_M \,.
\end{eqnarray}
where $U_l$, $U_\nu$ and $U_0$ diagonalize on the left
respectively the charged lepton, $M_\nu$ and $M_D$.
The mixing matrix $V_M$ is here defined to verify
the equality $U_\nu\equiv U_0V_M$ and is such that
\begin{eqnarray}\label{eq:C}
V_M M_\nu^\Delta V_M^T &\equiv& {\cal C}=
M_D^\Delta V_0^\dagger \frac{1}{M_R} V_0^\star M_D^\Delta\,,
\end{eqnarray}
In the quark sector we introduce $Y_u$ and $Y_d$
to be the Yukawa matrices for up and
down sectors.
They can be diagonalized by
\begin{eqnarray}
Y_u=U_u Y_u^\Delta V_u^\dagger&\mbox{and}&Y_d=U_d Y_d^\Delta V_d^\dagger\,,
\end{eqnarray}
where the $Y^\Delta$ are diagonal and the $U$s and $V$s are unitary matrices.
Then the quark mixing matrix is given by
\begin{eqnarray}\label{CKMud}
U_{CKM}&=&U_u^\dagger U_d\,.
\end{eqnarray}
Notice that if there is a flavor symmetry that
constrains the Yukawa couplings in such a way that the diagonalizing
unitary matrices are fixed then the entries of $Y_l$ can still be
very different from the entries of $Y_d^T$. However both Yukawa
matrices are diagonalized by the same mixing matrices.
This is exactly the case in the presence of an
$A_4$ discrete flavor symmetry dinamically broken into $Z_3$
\cite{Morisi:2007ft,Altarelli:2005yp}
and can be partially true in the case of $S_3$ softly broken into
$S_2$ \cite{Morisi:2005fy}.
In this case we have
$$Y_l \approx Y_d^T \rightarrow U_l \simeq V_d^\star\,.$$
In the same way, if we call $Y_\nu$ the Yukawa coupling
that will generate the Dirac neutrino
mass matrix $M_D$, we have also the relation
\begin{eqnarray}
Y_\nu \approx Y_u^T &\rightarrow& U_0 \simeq V_u^\star\,.
\end{eqnarray}
This relation, together with the previous one, implies
$$U_{PMNS} \simeq V_d^T V_u^\star V_M\,.$$
If the Yukawa matrices are diagonalized by similar matrix on the
left and on the right, for example in minimal renormalizable
$SO(10)$ with only small contributions from the antisymmetric
representations such as ${\bf 120}$ or more important in models
where the diagonalization is strongly constrained by the flavor
symmetry, the previous relationship translates
into a relation between $U_{PMNS}$, $U_{CKM}$ and $V_M$.
In fact we have
\begin{eqnarray}\label{VM}
Y_u   \simeq Y_u^T \rightarrow V_u^\star = U_u &\mbox{and}&
Y_d   \simeq Y_d^T \rightarrow V_d^\star = U_d\,. \nonumber
\end{eqnarray}
Finally we get that $V_M$ satisfies eq. (\ref{eq:fund}).
The form of $V_M$ can be obtained under some assumptions about the
flavor structure of the theory.
Some flavor models give for example a correlation $V_M$ with $(V_M)_{13}=0$.
As a consequence of the from of the non trivial Quark-Lepton 
complementarity there are some predictions for the model. 
For example the prediction for $\theta_{13}^{PMNS}$ of
sec. {\bf\ref{sec:PMNS}} \cite{Chauhan:2006im}
and the correlations between CP violating phases and the mixing angle
$\theta_{12}$ of sec. {\bf\ref{sec:CP}} \cite{Picariello:2006sp}.

\subsection{$\tilde M_D$ in non trivial Quark-Lepton complementarity}
\label{sec:MD}
Let us investigate the value of Dirac neutrino mass matrix $\tilde M_D$
in the base where right-handed Majorana neutrino mass matrix,
 charged leptons mass matrix and weak gauge interactions
 are diagonal.
We define the unitary matrix $V_R$ by the diagonalization of $M_R$
\begin{eqnarray}\label{eq:MRdiag}
V_R M_R^\Delta V_R^T = M_R\,.
\end{eqnarray}
and we obtain
\begin{eqnarray}
\tilde M_D&=&U_l^\dagger M_D V_R^\star\,.
\end{eqnarray}
We want now to related this $\tilde M_D$ matrix to the $CKM$ mixing
matrix by using the previous result.
First of all we rewrite this matrix as
\begin{eqnarray}\label{eq:MtLM}
\tilde M_D &=&U_l^\dagger M_D V_R^\star
\nonumber\\&=&
U_l^\dagger U_0 M_D^\Delta V_0^\dagger V_R^\star\,.
\end{eqnarray}
Then we notice that the matrix $V_0^\dagger V_R^\star$ is related
via the ${\cal C}$ matrix to the diagonal low energy neutrino
mass matrix $m_{low}^\Delta$ and to $V_M$.
In fact we have
\begin{eqnarray}
V_M m_{low}^\Delta V_M^T &=& {\cal C}
\nonumber\\&=&
M_D^\Delta V_0^\dagger \frac{1}{M_R} V_0^\star M_D^\Delta
\nonumber\\&=&\label{eq:VM}
M_D^\Delta V_0^\dagger V_R^\star
\frac{1}{M_R^\Delta} V_R^T V_0^\star M_D^\Delta
\end{eqnarray}
where we used the inverse of eq. (\ref{eq:MRdiag})
\begin{eqnarray}
V_R^\star \frac{1}{M_R^\Delta} V_R^\dagger = \frac{1}{M_R}\,.
\end{eqnarray}
We multiple on the left and on the right both sides of eq. (\ref{eq:VM})
by $1/M_D^\Delta$ and we get
\begin{eqnarray}\label{eq:V0VR}
V_0^\dagger V_R^\star\frac{1}{M_R^\Delta} V_R^T V_0^\star
&=&
\frac{1}{M_D^\Delta} V_M m_{low}^\Delta V_M^T \frac{1}{M_D^\Delta}\,.
\end{eqnarray}
Once we computed the $V_0^\dagger V_R^\star$ matrix form eq. (\ref{eq:V0VR}),
by using eq. (\ref{eq:MtLM}), we get
\begin{eqnarray}
\tilde M_D &=&
U_l^\dagger U_0 M_D^\Delta V_0^\dagger V_R^\star
\nonumber\\&=&
U_{PMNS} V_M^\dagger M_D^\Delta V_0^\dagger V_R^\star
\nonumber\\
&=&\label{eq:Our}
\Omega^\dagger U_{CKM}^\dagger M_D^\Delta V_0^\dagger V_R^\star\,,
\end{eqnarray}
where in the last line we used the relations in eq.
(\ref{eq:PMNS}) and (\ref{eq:fund}).

\subsection{{\em Fully} determination of $V_0^\dagger V_R^\star$
 and $M_R^\Delta$}
Eq. (\ref{eq:Our}) is the equivalent of the general 
formula \cite{Casas:2001sr} in presence of non trivial Quark-Lepton
complementarity.
We observe that the main modification is the presence of $U_{CKM}^\dagger$
instead of $U_{PMNS}$ thanks to the fact the these matrices
are related each other through $V_M$ as shown in the relation
of eq. (\ref{eq:fund}).
Let us now compute the $V_0^\dagger V_R^\star$ matrix in 
a general scenario.
\\
In the following we use the experimental constraint from
\cite{Chauhan:2006im} that says us that $(V_M)_{13}$ is zero
and the allowed ranges for $\theta_{12}^{V_M}$
and $\theta_{23}^{V_M}$ are \cite{Chauhan:2006im}
\begin{eqnarray}
	\tan^2\theta_{12}^{V_M}\in[0.3,1.0]
&\quad\mbox{ and }\quad&
	\tan^2\theta_{23}^{V_M}\in[0.5,1.4]
	\,.
\end{eqnarray}
Let us denote with $m_i$ the complex low energy neutrino masses
obtained after the see-saw ($m_{low}^\Delta=\{m_1,m_2,m_3\}$),
and with $M_i$ the eigenvalues of the Dirac neutrino mass matrix
$M_D$ ($M_D^\Delta=\{M_1,M_2,M_3\}$).
We get
{
\begin{eqnarray}\label{eq:V0VRfull}
V_0^\dagger V_R^\star\frac{1}{M_R^\Delta} V_R^\dagger V_0^\star
&=&
\left(
\begin{array}{ccc}
\frac{(m_1 c^2_{12}+m_2 s^2_{12})}{M_1^2}&
\frac{-(m_1-m_2)c_{12}c_{23}s_{12}}{M_1 M_2}&
\frac{(m_1-m_2)c_{12}s_{12}s_{23}}{M_1 M_3}
\\
\frac{(m_2-m_1)c_{12}c_{23}s_{12}}{M_1 M_2}&
\frac{(m_1s^2_{12}c^2_{23}+ m_2c^2_{12}c^2_{23}+ m_3s^2_{23})}{M_2^2}&
\frac{s_{23}c_{23}(m_3-m_2c^2_{12}-m_1s^2_{12})}{M_2 M_3}
\\
\frac{(m_1-m_2)c_{12}s_{12}s_{23}}{M_1 M_3}&
\frac{s_{23}c_{23}(m_3-m_2c^2_{12}-m_1s^2_{12})}{M_2 M_3}&
\frac{s^2_{23}(m_1s^2_{12}+ m_2c^2_{12})+ m_3c^2_{23}}{M_3^2}
\\
\end{array}
\right)\,.
\nonumber
\end{eqnarray}
}
\begin{eqnarray}
\
\end{eqnarray}
Eq. (\ref{eq:V0VRfull}) is general and must be specified depending
on the explicit form of $V_M$.
We have three cases \cite{Picariello:2007yn}:
\begin{enumerate}
\item{hierarchical Dirac neutrino eigenvalues 
(very hierarchical right-handed neutrino masses,
$V_0^\dagger V_R^\star\simeq I$) where we get the usual ratios
$$BR(\mu\rightarrow e\gamma):BR(\tau\rightarrow e\gamma):BR(\tau\rightarrow \mu\gamma)
  =\lambda^6:\lambda^4:1 \propto M_3^4 \lambda^4 \hat L\,;$$
}
\item{degenerate Dirac neutrino eigenvalues, with non degenerate
low energy neutrino masses
(the hierarchy of the right-handed neutrino masses is close to the
one of the low energy spectrum, $V_0^\dagger V_R^\star \simeq V_M$)
where we get
$$BR(\mu\rightarrow e\gamma)= \tan^2\theta_{23}^{V_M}
 BR(\tau\rightarrow e\gamma)
= f(\theta_{12}^{V_M},\theta_{23}^{V_M}) BR(\tau\rightarrow \mu\gamma) \propto M_3^4 \hat L$$
with $f(\theta_{12}^{V_M},\theta_{23}^{V_M})$ of order one \cite{Picariello:2007yn};
}
\item{degenerate Dirac neutrino eigenvalues and low energy neutrino
spectrum (right-handed neutrino close to be degenerate,
$V_0^\dagger V_R^\star\simeq I$) where we have
$$BR(\mu\rightarrow e\gamma):BR(\tau\rightarrow e\gamma):BR(\tau\rightarrow \mu\gamma)
=1:\lambda^4:\lambda^2 \propto M_3^4\lambda^{10} \hat L\,.$$
}
\end{enumerate}
\subsection*{Acknowledgments}
We thanks Jorge C. Rom\~ao for enlighting
discussion about flavor violating processes in supersymmetry.
We acknowledge the MEC-INFN grant, and
the Funda\c{c}\~{a}o para a Ci\^{e}ncia e a Tecnologia for the grant SFRH/BPD/25019/2005.

\def\OLD{
\bibitem{SNO}
B.~Aharmim {\it et al.}  [SNO Collaboration],
Phys.\ Rev.\ C {\bf 72} (2005) 055502
\bibitem{Krauss:2006qq}
C.~B.~Krauss  [SNO Collaboration],
J.\ Phys.\ Conf.\ Ser.\  {\bf 39} (2006) 275.
\bibitem{SKatm}
[Super-Kamiokande Collaboration],
arXiv:hep-ex/0607059.
\bibitem{SKatm2}
J.~Hosaka {\it et al.}  [Super-Kamiokande Collaboration],
Phys.\ Rev.\ D {\bf 74} (2006) 032002
\bibitem{SKsolar}
J.~Hosaka {\it et al.}  [Super-Kamkiokande Collaboration],
Phys.\ Rev.\ D {\bf 73} (2006) 112001
\bibitem{GNO}
M.~Altmann {\it et al.}  [GNO Collaboration],
Phys.\ Lett.\ B {\bf 616} (2005) 174
%
\bibitem{GALLEX}
W.~Hampel {\it et al.}  [GALLEX Collaboration],
Phys.\ Lett.\ B {\bf 447} (1999) 127.
%
\bibitem{HOMESTAKE} 
B.~T.~Cleveland {\it et al.} [HOMESTAKE],
Astrophys.\ J.\  {\bf 496} (1998) 505.
\bibitem{SAGE}
J.N. Abdurashitov et al. [SAGE Collaboration] Phys. Rev. Lett. 83(23) (1999)4686.
\bibitem{KamLAND} 
T.~Araki {\it et al.}  [KamLAND Collaboration],
Phys.\ Rev.\ Lett.\  {\bf 94} (2005) 081801
\bibitem{CHOOZ}
M.~Apollonio {\it et al.}  [CHOOZ Collaboration],
Phys.\ Lett.\ B {\bf 466} (1999) 415
\bibitem{PaloVerde} 
Y.~F.~Wang  [Palo Verde Collaboration], 
Int.\ J.\ Mod.\ Phys.\ A {\bf 16S1B}, 739 (2001); 
\bibitem{PaloVerde2}
F.~Boehm {\it et al.} [Palo Verde Collaboration], 
Phys.\ Rev.\ D {\bf 64}, 112001 (2001). 
\bibitem{MINOS}
[MINOS Collaboration],
arXiv:hep-ex/0607088.
}
\def\OLD{
\bibitem{Chen:2006rk}
B.~L.~Chen, H.~L.~Ge, C.~Giunti and Q.~Y.~Liu,
  arXiv:hep-ph/0605195.
\bibitem{Robertson:2006pk}
  R.~G.~H.~Robertson,
  Prog.\ Part.\ Nucl.\ Phys.\  {\bf 57} (2006) 90
  [arXiv:nucl-ex/0602005].
\bibitem{McDonald:2006qf}
  A.~B.~McDonald,
  J.\ Phys.\ Conf.\ Ser.\  {\bf 39} (2006) 211.
\bibitem{Petcov:2006gy}
  S.~T.~Petcov and T.~Schwetz,
  arXiv:hep-ph/0607155.

\bibitem{Barger:2006vy}
  V.~Barger, M.~Dierckxsens, M.~Diwan, P.~Huber, C.~Lewis, D.~Marfatia and B.~Viren,
  Phys.\ Rev.\ D {\bf 74} (2006) 073004
  [arXiv:hep-ph/0607177].

\bibitem{unknown:2006mn}
    [Double Chooz Collaboration],
  arXiv:hep-ex/0606025.

\bibitem{Bernabeu:2006az}
  J.~Bernabeu and C.~Espinoza,
  arXiv:hep-ph/0605132.

\bibitem{McFarland:2006pz}
  K.~S.~McFarland  [MINERvA Collaboration],
  arXiv:physics/0605088.

\bibitem{Savvinov:2006pb}
  N.~Savvinov  [OPERA Collaboration],
  arXiv:hep-ex/0602010.

\bibitem{Decowski:2006zg}
  M.~P.~Decowski  [KamLAND Collaboration],
  Acta Phys.\ Polon.\ B {\bf 37} (2006) 245.

\bibitem{Ochoa-Ricoux:2006qn}
  J.~P.~Ochoa-Ricoux  [MINOS Collaboration],
  Prog.\ Part.\ Nucl.\ Phys.\  {\bf 57} (2006) 147.

\bibitem{Kraus:2006qp}
  C.~Kraus  [SNO+ Collaboration],
  Prog.\ Part.\ Nucl.\ Phys.\  {\bf 57} (2006) 150.

\bibitem{Bhattacharya:2006ri}
  S.~Bhattacharya  [INO Collaboration],
  Prog.\ Part.\ Nucl.\ Phys.\  {\bf 57} (2006) 299.

\bibitem{Ciesielski:2006ie}
  R.~Ciesielski  [OPERA Collaboration],
  Acta Phys.\ Polon.\ B {\bf 37} (2006) 1237.

\bibitem{Vignoli:2006jz}
  C.~Vignoli, D.~Barni, J.~M.~Disdier, D.~Rampoldi and G.~Passardi  [ICARUS
                  Collaboration],
  AIP Conf.\ Proc.\  {\bf 823} (2006) 1643.


\bibitem{Aguilar:2006rm}
  J.~A.~Aguilar {\it et al.}  [ANTARES Collaboration],
  arXiv:astro-ph/0606229.

\bibitem{Bouchta:2006rt}
  A.~Bouchta  [IceCube Collaboration],
  arXiv:astro-ph/0606235.

\bibitem{Ahn:2006zz}
  M.~H.~Ahn {\it et al.}  [K2K Collaboration],
  arXiv:hep-ex/0606032.

\bibitem{Balata:2006ue}
  M.~Balata {\it et al.}  [Borexino Collaboration],
  arXiv:hep-ex/0601035.

\bibitem{Aggouras:2006mm}
  G.~Aggouras {\it et al.}  [NESTOR Collaboration],
  Nucl.\ Phys.\ Proc.\ Suppl.\  {\bf 151} (2006) 279.

\bibitem{Broggini:2006rm}
  C.~Broggini  [LUNA Collaboration],
  Prog.\ Part.\ Nucl.\ Phys.\  {\bf 57} (2006) 343.
}
\def\OLD{
\bibitem{Balantekin:2004hi}
A.~B.~Balantekin, V.~Barger, D.~Marfatia, S.~Pakvasa and H.~Yuksel, 
arXiv:hep-ph/0405019. 
\bibitem{Oberauer:2004ji} 
L.~Oberauer, 
Mod.\ Phys.\ Lett.\ A {\bf 19} (2004) 337 
\bibitem{Rodejohann:2006ek}
  W.~Rodejohann,
  Phys.\ Scripta {\bf T127} (2006) 35.
\bibitem{Gonzalez-Garcia:2006wm}
  M.~C.~Gonzalez-Garcia, M.~Maltoni and J.~Rojo,
  arXiv:astro-ph/0608107.
\bibitem{Giunti:2006fr}
  C.~Giunti,
  arXiv:hep-ph/0608070.
\bibitem{Valle:2006vb}
  J.~W.~F.~Valle,
  arXiv:hep-ph/0608101.
\bibitem{Fogli:2006jk}
  G.~L.~Fogli, E.~Lisi, A.~Mirizzi, D.~Montanino and P.~D.~Serpico,
  arXiv:hep-ph/0608321.
\bibitem{Bandyopadhyay:2006jn}
  A.~Bandyopadhyay, S.~Choubey, S.~Goswami and S.~T.~Petcov,
  arXiv:hep-ph/0608323.
\bibitem{Bilenky:2006sn}
  S.~M.~Bilenky,
  arXiv:hep-ph/0607317.
\bibitem{Messier:2006yg}
  M.~D.~Messier,
  eConf {\bf C060409} (2006) 018
  [arXiv:hep-ex/0606013].
\bibitem{Strumia:2006db}
  A.~Strumia and F.~Vissani,
  arXiv:hep-ph/0606054.
\bibitem{Schwetz:2006dh}
  T.~Schwetz,
  Phys.\ Scripta {\bf T127} (2006) 1
  [arXiv:hep-ph/0606060].
\bibitem{Fukugita:2006rm}
  M.~Fukugita, K.~Ichikawa, M.~Kawasaki and O.~Lahav,
  Phys.\ Rev.\ D {\bf 74} (2006) 027302
  [arXiv:astro-ph/0605362].
}
\def\Texture{
\bibitem{Dev:2006qe}
  S.~Dev, S.~Kumar, S.~Verma and S.~Gupta,
  arXiv:hep-ph/0612102.
\bibitem{Dev:2006xu}
  S.~Dev, S.~Kumar, S.~Verma and S.~Gupta,
  arXiv:hep-ph/0611313.
\bibitem{Albright:2006cw}
  C.~H.~Albright and M.~C.~Chen,
  Phys.\ Rev.\  D {\bf 74} (2006) 113006
  [arXiv:hep-ph/0608137].
\bibitem{Dev:2006if}
  S.~Dev and S.~Kumar,
  arXiv:hep-ph/0607048.
\bibitem{Merle:2006du}
  A.~Merle and W.~Rodejohann,
  Phys.\ Rev.\  D {\bf 73} (2006) 073012
  [arXiv:hep-ph/0603111].
\bibitem{Singh:2006dr}
  N.~N.~Singh, M.~Rajkhowa and A.~Borah,
  arXiv:hep-ph/0603189.
\bibitem{Singh:2006uw}
  N.~N.~Singh, M.~Rajkhowa and A.~Borah,
  J.\ Phys.\ G {\bf 34} (2007) 345
  [arXiv:hep-ph/0603154].
}
\def\COSMO{
\bibitem{Joaquim:2003pn}
  F.~R.~Joaquim,
  Phys.\ Rev.\  D {\bf 68} (2003) 033019
  [arXiv:hep-ph/0304276].
\bibitem{MacTavish:2005yk}
  C.~J.~MacTavish {\it et al.},
  Astrophys.\ J.\  {\bf 647} (2006) 799
  [arXiv:astro-ph/0507503].
\bibitem{Hannestad:2005ey}
  S.~Hannestad,
  Prog.\ Part.\ Nucl.\ Phys.\  {\bf 57} (2006) 309
  [arXiv:astro-ph/0511595].
\bibitem{Hannestad:2005gj}
  S.~Hannestad,
  Phys.\ Rev.\ Lett.\  {\bf 95} (2005) 221301
  [arXiv:astro-ph/0505551].
\bibitem{Xia:2006wd}
  J.~Q.~Xia, G.~B.~Zhao and X.~Zhang,
  Phys.\ Rev.\  D {\bf 75} (2007) 103505
\bibitem{Kirilova:2006wh}
  D.~P.~Kirilova and M.~P.~Panayotova,
  JCAP {\bf 0612} (2006) 014
}

\end{document}